# Growth Kinetics of Ion Beam Sputtered Al-thin films by Dynamic Scaling Theory

Neha Sharma*, K. Prabakar, S. Dash and A. K. Tyagi

Material Science Group,

Indira Gandhi Center for Atomic Research,

Kalpakkam, TN, 603102

**Abstract**

This paper reports the study of growth kinetics of ion beam sputtered aluminum thin films. Dynamic scaling theory was used to derive the kinetics from AFM measurements. AFM imaging revealed that surface incorporates distinctly different morphologies. Variation in deposition times resulted in such distinctiveness. The growth governing static ($\alpha$) as well as dynamic ($\beta$) scaling exponents have been determined. The exponent $\alpha$ decreased as the deposition time increased from 3 to 15 minutes. Consequently, the interfacial width ($\xi$) also decreased with critical length ($L_c$), accompanied with an increase in surface roughness. Surface diffusion becomes a major surface roughening phenomenon that occurs during deposition carried out over a short period of 3 minutes. Extension of deposition time to 15 minutes brought in bulk diffusion process to dominate which eventually led to smoothening of a continuous film.

*Corresponding author: Neha Sharma
neha@igcar.gov.in
Indira Gandhi Centre for Atomic Research

1. Introduction

Thin films technology has been evolved as a potential solid state material processing technique. Thus qualitative understanding of thin film growth is essential in case of each deposition technique to optimize film microstructure. Three consecutive steps are encountered in deposition of a thin film. Initially the formation of deposition species (atomic, molecular or ionic species) takes place. This is followed by their transport from source to substrate. The final step points to their condensation onto the substrate and the subsequent film growth. Based on these expects, three distinct growth mechanisms have been proposed[1]. Volmer-Weber growth (or 3D island growth) of small clusters nucleating directly on the substrate surface and then growing together to form 3D-islands. These 3D-islands in turn coalesce to form a continuous film. Frank-Vander Merwe growth (2D layer-by-layer growth) causes layers to form and grow on to the substrate. Stranski-Krastanova growth (mixed mode growth) follows layer-by-layer mode to start with and after forming one or two monolayers, layer growth gives way to 3D island mode of growth[1]. Depending upon various growth conditions and materials, different growth mechanisms are encountered. These are specific to deposition techniques and process conditions. In the light of these facts, it is essential to define specific parameters for pin pointing growth mechanisms encountered in varied deposition techniques and process conditions.

A possible and reliable way of assessing the growth mechanism of thin films involves mathematical analysis of surface topography within the frame work of dynamic scaling theory (DST)[2]. This provides a means to acquire the information about the effects of certain surface phenomenon like plastic flow, condensation, evaporation and diffusion on the thin film growth. Therefore the role of such physical processes on the microstructural properties such as roughness and conformality is easily addressed.

In 1985, Family and Vicsek[3] propounded dynamic scaling theory to analyze the behavior of growing surfaces by assuming that these are self-affine. According to the conventional dynamic scaling theory, the scaling behavior can be represented by the Family-Vicsek scaling relation[3].

$$\xi(L,t) = L^\alpha f(t/L^z) \qquad (1)$$

Where

$$\xi(L,t) \sim t^\beta \quad for \quad t/L^z \ll 1$$

and

$$\xi(L,t) \sim L^\alpha \quad for \quad t/L^z \gg 1$$

Here, $\xi$ is the interface width, L is the length scale over which the roughness is measured, t is the time of growth. $\alpha$ and $\beta$ are static and dynamic scaling exponents, respectively. z equals $\alpha/\beta$. Such a scaling behavior holds well for thin films synthesized by techniques like evaporation[4], sputtering[5], thermal chemical vapor deposition (CVD)[6] and PECVD[7].

However, the scaling behavior defined by Eq. (1) is not satisfied in other experimental systems[8, 9]. There is always an anomalous scaling present in every experimental system which generally causes the appearance of distinct values of the scaling exponents, $\alpha$ and $\beta$, which depends upon the scale of measurement[10].

In this work, we are motivated to develop a general understanding of the fundamental processes controlling the growth of the Al-thin film by Ion beam sputter deposition system (IBSD). Al-thin film is a promising candidate for microelectronic device fabrication due to its low resistivity and high compatibility with silicon. For this, we have determined the scaling exponents of Al-thin films deposited on Si(100) by IBSD. The purpose of this study is to understand the growth kinetics of IBSD grown Al-thin films as a function of deposition time and scale of measurement.

## 2. Theoretical Background

Conventional DST considers development of self-affine surfaces to be associated with scaling relations among surface roughness (defined as root mean square (rms) of surface height H(r) and denoted by "δ"), deposition time (t) and scale of measurement (L). As mentioned earlier, DST proposes that the behavior of the interface width follows the trends propounded by Family–Vicsek relationship[3] given by equation (1). This exhibits two distinct asymptotic scaling behavior for the interface width ξ:

$$\xi(L,t) \sim t^\beta \quad for \quad t <<$$
$$and \quad \xi(L,t) \sim L^\alpha \quad for \quad t >>$$

Therefore, during initial growth stages, the interface width increases with deposition time 't' at a rate of the growth exponent β until a saturation value of $\xi_L$ is reached. After attaining this value, the interface width becomes a function of scale of measurement 'L' through the growth exponent α. At some critical length $L_c$, the interface width saturates and becomes equal to rms value of roughness 'δ' for all $L > L_c$.

A quantitative representation on the height variation and lateral correlation is provided by the "autocovariance function G(|r|)" [11]. G(|r|) at different length scales provides a quantitative description of correlation among heights at different points on a surface as a function of their separation "r". The interface width as a function of length scale is related to this autocorrelation function in the following way[12, 13]:

$$\xi_L^2 = \left(\frac{1}{L^2}\right) \int_0^L [\delta^2 - G(|r|)] r \, dr \qquad (2)$$

The autocorrelation function for a self-affine surface with spatial scaling exponent α can be approximated as:

$$G(|r|) \approx \begin{cases} \delta^2 \left[1 - \frac{\alpha+1}{2}\left(\frac{r}{L_c}\right)^{2\alpha}\right], & \text{for } r \leq L_c \\ 0, & \text{for } r > L_c \end{cases} \quad (3)$$

The Fourier transform of the equation (3) yields power spectral density function (PSD) $g(|q|)$. This is a very useful function as several fundamental aspects of a rough surface can be formulated in terms PSD. Mathematically PSD is linked to autocovariance function as mentioned below.

$$g(|q|) = \mathscr{F}[G(|r|)]$$

where $\mathscr{F}$ is the two-dimensional Fourier transform operator. The PSD assumes the forms:

$$g(|q|) \approx \begin{cases} \frac{\alpha}{\pi}\delta^2 L_c^2 & \text{for} & |q| < 1/L_c \\ \frac{\alpha}{\pi}\frac{\delta^2}{L_c^{2\alpha}} q^{-2(\alpha+d)} & \text{for} & |q| \geq 1/L_c \end{cases} \quad (4)$$

where d, in our case, represents line scan direction and equals to '1'.

Thin film growth via a surface-vapor interaction takes place by the stochastic addition or removal of atoms with no lateral transport occurring on the surface. The scaling of growing self-affine surface arises from the competition between roughening and various smoothening mechanisms. By combining the smoothening mechanisms with the stochastic roughening, a kinetic rate expression can be written in reciprocal space to account for surface growth [14]:

$$\frac{\partial h(|q|,t)}{\partial t} \propto -c_n |q|^n h(q,t) + \acute{\eta}(|q|,t) \qquad (n = 1,2,3,4) \qquad (5)$$

Here $h(|q|, t)$ represents the radial average of the Fourier transformed surface height denoted by $H(|r|, t)$. First term on the right hand side represents the smoothening mechanism described by Herring et al.[15]. The second term is the stochastic noise term in reciprocal space that describes random arrival of the depositing species. The solution to equation (5) is obtained as radially averaged power spectral density[14]:

$$g(|q|, t) = <h(|q|,t)^2> = \Omega \frac{1 - \exp(-2c_n|q|^n t)}{c_n |q|^n} \qquad (6)$$

In this equation $\Omega$ is proportional to flux, t is the deposition time, and $c_n$ is a constant which characterizes the specific lateral mass transport mechanism indicated by "n".

We have used a fitting function similar to equation (6) to model our data in the manner similar to one used by William M. Tong et.al.[16]. This is expressed as:

$$g(|q|, t) \propto \Omega \frac{\exp(2\sum \chi_n |q|^n t) - 1}{\sum \chi_n |q|^n} \qquad (7)$$

Where the q-coefficients $\chi_n$ are simply fitting parameters and n takes the values 1,2,3,4.

The q-coefficients are allowed to acquire either positive or negative values which will determine whether the process will be smoothening and roughening one. According to Herring [15], four smoothening mechanisms are plastic flow (n = 1), evaporation-recondensation (n = 2), bulk diffusion (n = 3) and surface diffusion (n = 4).

3. **Experiment**

Thin films analyzed in this study were grown by IBSD on Si (100) substrates for different time durations. IBSD system essentially consists of a main deposition chamber and a load lock chamber with a substrate transfer rod. Main deposition chamber is equipped with quartz crystal monitor, residual gas analyzer, substrate heater, substrate rotator and other high vacuum measuring gauges. Prior to deposition, substrates were cleaned by RCA-1 process[17].

Deposition was carried out at room temperature, at a base pressure of $4 \times 10^{-6}$ mbar with working pressure at $2 \times 10^{-4}$ mbar. Thin films were grown for 3, 5, 8 and 15 minutes. During deposition, the metal atom flux was supplied by sputtering Al target with an inert ion beam of Ar+ ions having energy of 500 eV.

The surface morphology was analyzed using an atomic force microscope (AFM) (Ntegra Prima of M/s NT-MDT, Russia) in semi-contact mode. A Si-tip of radius 35 nm was used to scan 1x1 $\mu m^2$ area over the film surface. Roughness of the film was calculated as root mean square of the surface heights acquired at different points on the surface. In order to improve the statistics and representivity, images and roughness line profile were acquired at several points on the specimen surface.

The static scaling exponent "α" was calculated by plotting the logarithm of interfacial width (ξ) against logarithm of scan length (L) for each of film. The dynamic scaling exponent "β" was obtained from the slope of the curve generated by log-log plot of rms surface roughness (δ) vs. deposition time (t). Scaling properties of the films were further studied using an autocorrelation function ($G(|r|)$) and power spectral density PSD expressed as g(q) to obtain information about different roughening and smoothening phenomena occurring on the film surface. G($|r|$) and g($|q|$) curves were obtained for different films grown for different deposition times using the procedure described earlier. Distinctly different trends were observed for films deposited over different deposition durations.

## 4. Results and Discussion

In figure. 1, the evolution of surface morphology is shown for the films grown over different deposition times of 3, 5, 8 and 15 minutes. It is inferred from the figure. 1(A) that small nuclei are nucleated at the substrate surface after a deposition of 3 minutes. This signals onset of thermal equilibrium between deposition species and substrate surface. As the deposition time increases, these tiny nuclei grow in number as well as in size forming bigger clusters as shown in figure 1(B). In the figure 1(C), a discontinuous morphology of the film is observed due to coalescence of the bigger clusters. Such clusters coalescence is a consequence of reduction in surface area which leads to full coverage of the substrate surface that leaves behind very few uncovered channels. With further increase in deposition time, secondary nucleation takes place on uncovered channels eventually forming a continuous film as shown in figure 1(D).

This can also be appreciated well from the figure 2 where the representative line profiles along the images are presented. In these samples, grown over different deposition times, there are two length scales: the first one is characterized by the height fluctuations among different nuclei, clusters or islands and the second one is defined by small aggregates of these nuclei and clusters. It is possible to observe initial stages of the development of surface topography from figure 2(A) in the form of tiny nuclei. These nuclei exhibit peaks constituted of small height amplitudes. As the deposition time is increased, the amplitude of height oscillations and corresponding spatial coverage increase as inferred from comparatively larger and broader peaks in the line profile of figure 2(B). Line profiles in figure 2(A) and (B) together support the fact that small nuclei nucleated at early deposition stages grow with time to form larger clusters. These clusters then agglomerate to form islands as is shown by line profile in figure 2(C). This figure shows large spatial coverages concomitant with large height oscillations. This phenomenon leads to deposition over larger area of substrate surface with few channels left out to be filled by secondary nucleation process. Finally in figure 2(D) the appearance of smaller and rarely equal height amplitudes are seen. The complete spatial coverage of substrate surface following above hierarchy of events is indicative of deposition of a smooth continuous film.

An important fact is observed while calculating the static scaling exponent or roughness exponent "α". According to dynamic scaling theory as introduced briefly in Sec. 2, the roughness exponent can be calculated by plotting log ($\xi$) vs. log (L) as shown in figure 3. At small length scales, $\xi$ increases with length scale and becomes saturated ($\xi_L$) after a certain critical length $L_c$. This saturated value of $\xi_L$ is equal to the rms roughness '$\delta$' of the surface under investigation. Figure 3(A) shows that an α = 0.61 value governs the evolution of surface roughness until a saturated value of rms surface roughness, $\delta$ = 0.269nm, is achieved at a critical length $L_C$ = 631 nm, for a 3 minutes deposition. As the deposition time increases, α-values begin to decrease with smaller critical lengths and '$\delta$' increases to 0.550nm for 8 minutes deposition until film achieves larger surface coverage on the substrate barring few uncovered channels. Once the film becomes continuous and smooth after a deposition of 15 minutes, α becomes lowest with a value of 0.38 concomitant with low '$\delta$' value that equals to 0.398nm. This is shown in figure 3(B), 3(C) and 3(D), respectively. The corresponding values of α, $L_C$, and $\delta$ are listed in table 1. These α-values match with those obtained from the slope of the linear region of PSD plot shown in figure 8.

The dynamic scaling exponent 'β' acquires two different values as shown in figure 5. This behaviour of 'β' suggests to divide the growth in to two regimes. One is before full coverage of the substrate surface governed by $\beta_1 = 0.73$ and other is after full coverage of substrate which is governed by $\beta_2 = -0.52$.

Figure 6 represents further information about both height variation and lateral correlation predicted from autocovariance function $G(|r|)$ on all the samples. It provides information on height correlation across lateral direction. Fourier transform of $G(|r|)$, given by equation(4), provides the function $g(|q|)$ or PSD which is shown in figure 7. When this g(q) is fitted to equation (7), it identifies the growth governing phenomena from material related events like plastic flow, evaporation-condensation, bulk and surface diffusions. Either one or more than one phenomenon may dominate a particular film growth process under a given deposition condition.

Generally during the course of thin film deposition, adsorbed species are not in thermal equilibrium with the substrate initially and interact among themselves as well as with the substrate surface to achieve thermal equilibrium [1,20]. Few such common phenomenon being investigated in this study are plastic flow, evaporation-recondensation, bulk and surface diffusions. Different extents of these processes act together during the growth of a thin film. These phenomenon are represented by n = 1,2,3,4 respectively in equation (7). Figure 8 shows the result of fitting of experimental $g(|q|)$ for different deposition time with equation (7). It can easily be seen that this fitting function captures many of the important features of the experimental $g(|q|)$ or PSD functions. The sign of the parameters $\xi_n$, shown in table 2, decides whether the corresponding smoothening or roughening mechanism will be dominant. Although all the fits are not consistent but fits for t = 8 minutes and t = 15 minutes follow the same set of growth mechanisms. For sample grown for 3 minutes, primary roughening process is related to n = 2 and n = 4 where surface diffusion (n=4) prevails as roughening phenomenon. In case of smoothening phenomenon bulk diffusion corresponding to n = 3 plays a major role along with plastic flow of surface shown in figure 8(A). Figure 8(B) shows that as the deposition time is increased to 5 minutes, bulk diffusion corresponding to n = 3 becomes the leading roughening mechanism while surface diffusion with n = 4 smoothens the surface voluntarily. Thus, the vapor atoms reaching on the surface diffuse over it and interact among themselves to form clusters (as shown in figure 1(B)). As the deposition time increases further to 8 minutes, surface coverage increases by plastic flow (n=1), bulk diffusion (n=3) and surface diffusion (n=4). n = 2 corresponds to mechanism governed by evaporation and recondensation process to bring about roughening. Thus, the film covers a larger area on the substrate surface with few uncovered channels and holes being left out. This is indicated in figure 8(C). For a 15 minutes deposition schedule, shown in figure 8(D), the film becomes continuous with a lower surface roughness. In this case bulk diffusion behaviour corresponding to n = 3 plays a dominating role in smoothening of film surface by filling up of all the valleys and irregularities present on the surface. In this case both plastic flow and surface diffusion contribute very little. Once again evaporation and recondensation become major roughening mechanism. Present investigations, however, do not rule out existence of additional mechanisms responsible for smoothening and roughening.

## 5. Conclusion

In principle, the entire spatial and temporal dependence of a surface can be summarized by the static and dynamic scaling exponents ($\alpha$, $\beta$), the material-dependent surface rms height, and the critical length for scaling [$\delta(\tau)$, $L_c(\tau)$] measured at particular time ($\tau$). In this study, the static scaling exponent '$\alpha$' decreases as the deposition time increases. Consequently, the interfacial width becomes smaller which is associated with smaller critical lengths ($L_c$). '$\delta$' increases with deposition time until film achieves maximum coverage on the substrate surface and once the film becomes continuous, 'δ' achieves a lower value and film subsequently becomes smoother. The dynamic scaling exponent 'β' does not achieve a universal value.

To support above observations and after examining figure (8) it is concluded that fundamental mechanisms leading the growth of surface morphologies are different for different deposition times. For a 3 minutes deposition, surface gets rougher by evaporation and recondensation of the vapor atoms reaching on the surface along with substantial diffusion across lateral directions resulting in the formation of small nuclei. A 5 minutes deposition under the same conditions, forms 3D- clusters of adatoms due to the dominance of surface diffusion. Then for an 8 minutes deposition, bulk diffusion dominates to fill the valleys and other left over irregularities on the surface and covers a larger area on the substrate surface. Both plastic flow and bulk diffusion act as main smoothening phenomena. At longer deposition times of 15 minutes, same set of smoothening phenomena operates to achieve a continuous film with no uncovered channels left out on the surface.

**Acknowledgments**

The authors acknowledge, Dr. C. S. Sunder, Director, Material Science Group for encouragement.

**Figure captions:**

**Figure. 1** Representative 3D AFM images (1 X 1 µm$^2$) of the evolution of surface morphology for the films grown for different deposition times as, (A) 3 minutes, (B) 5 minutes, (C) 8 minutes and (D) 15 minutes.

**Figure. 2** Representative line profiles across the sample surface for different deposition times: (A) 3 minutes, (B) 5 minutes, (C) 8 minutes, (D) 15 minutes.

**Figure. 3** Plot of the interface width of the substrate surface as a function of length scale for several different growth times varied as (A) 3 minutes (B) 5 minutes (C) 8 minutes and (D) 15 minutes.

**Figure. 4** Comparison of α–exponents obtained from two different techniques for different deposition times.

**Figure. 5** Dynamic scaling constant 'β' governing the growth for different deposition times with different slopes

**Figure. 6** Autocovariance function for different deposition times depicting how heights are co-related at different points across lateral direction

**Figure.7** Spectral power density obtained corresponding to different deposition times an verifying critical lengths obtained for each deposition time.

**Figure. 8** Fitting of experimentally obtained g(q) with equation (7) for different deposition times indicating towards dominating roughening/smoothening phenomenons. The data are presented as log–log plots where (A) 3 minutes, (B) 5 minutes, (C) 8 minutes and (D) 15 minutes.

**Tables:**

| Deposition Time 't' (in min.) | α | $L_C$ (in nm) | δ (in nm) |
|---|---|---|---|
| 3 | 0.61 | 631 | 0.269 |
| 5 | 0.48 | 530 | 0.426 |
| 8 | 0.41 | 430 | 0.550 |
| 15 | 0.38 | 430 | 0.398 |

**Table-1 Comparison of α, $L_c$ and δ for different deposition times**

| Samples | $\Omega$ (X $10^3$) | t (in min.) | $\xi_1$ (X $10^3$) | $\xi_2$ (X $10^6$) | $\xi_3$ ($10^8$) | $\xi_4$ (X $10^{10}$) |
|---|---|---|---|---|---|---|
| A | 3.069 | 3 | -2.171 | 2.552 | -9.798 | 5.642 |
| B | 2.223 | 5 | -0.188 | -0.462 | 5.366 | -14.802 |
| C | 2.172 | 8 | -0.609 | 0.389 | -0.684 | -0.469 |
| D | 10.080 | 15 | -0.608 | 4.121 | -8.753 | -0.518 |

**Table-2 Comparison of fits of equation (7) for different deposition times**

**Figure. 1**

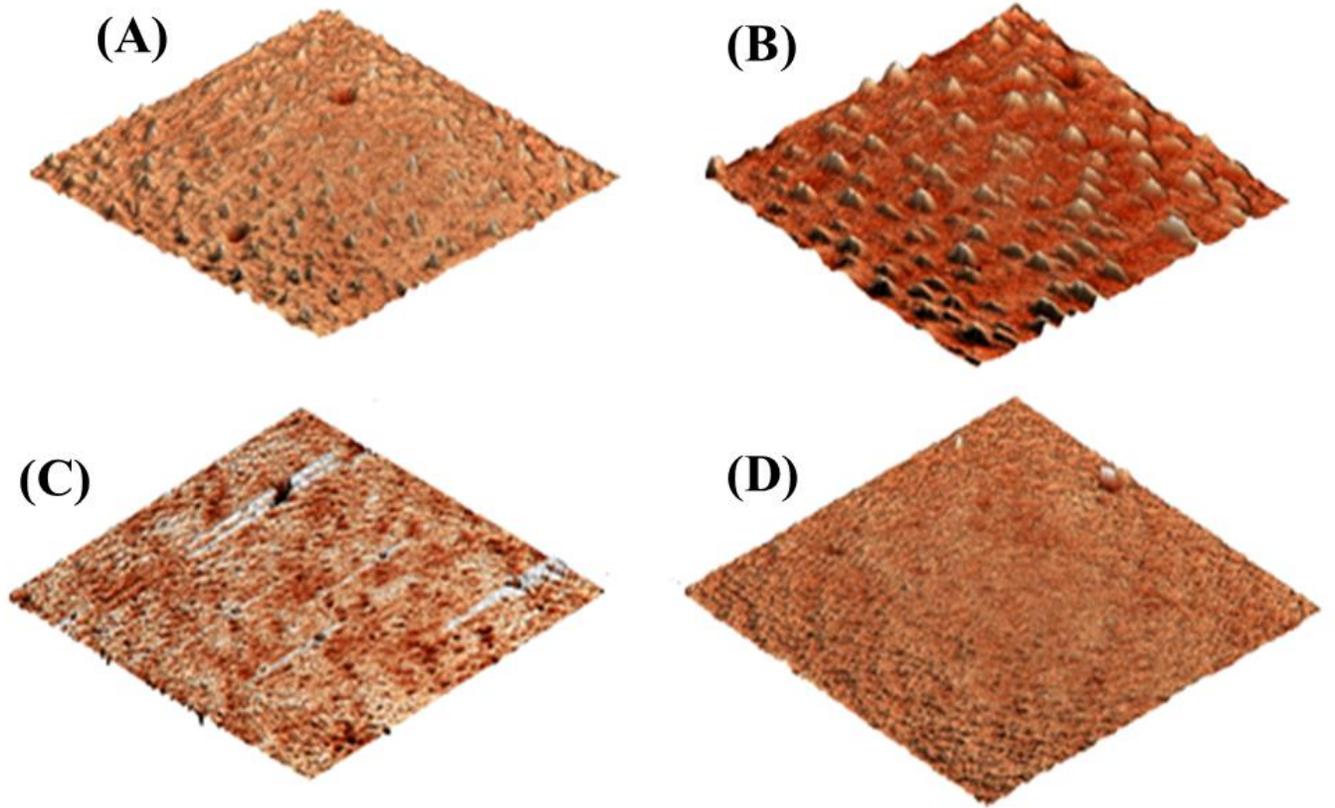

**Figure. 2**

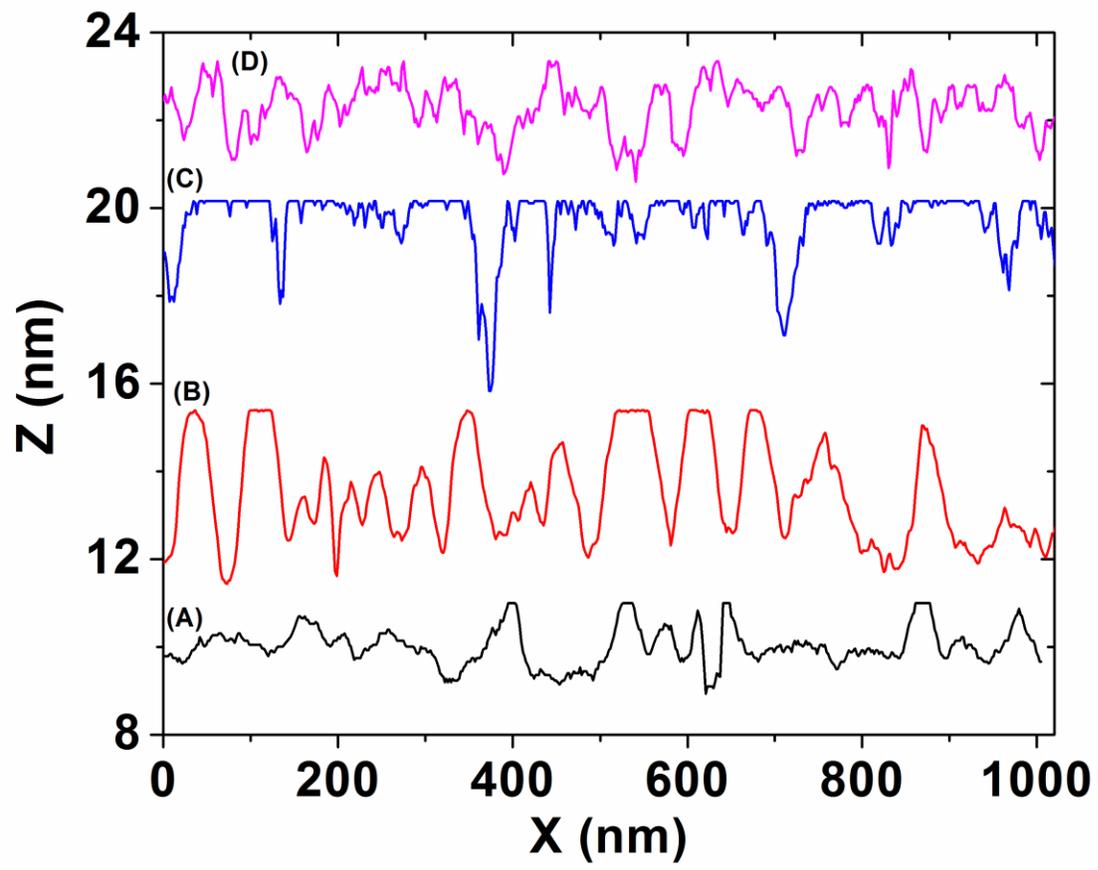

**Figure. 3**

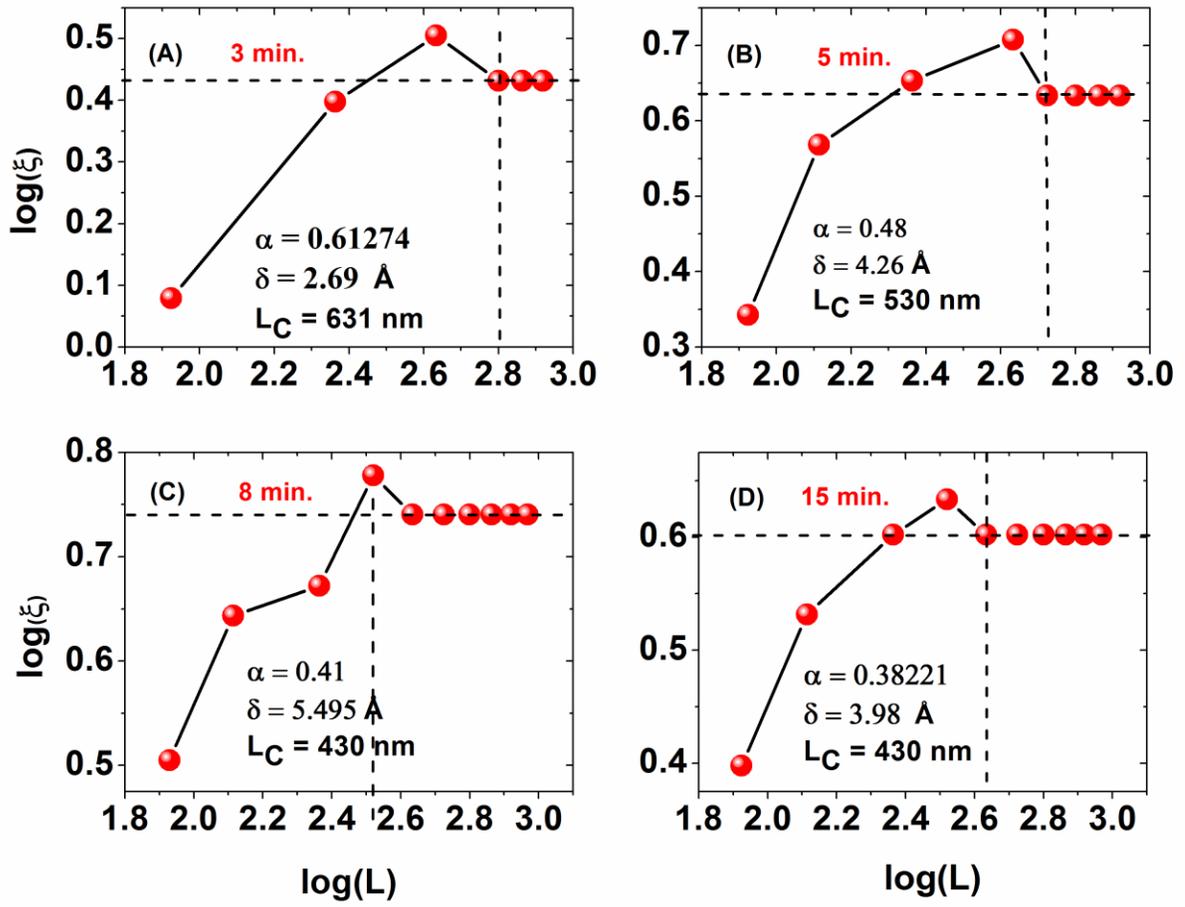

**Figure. 4**

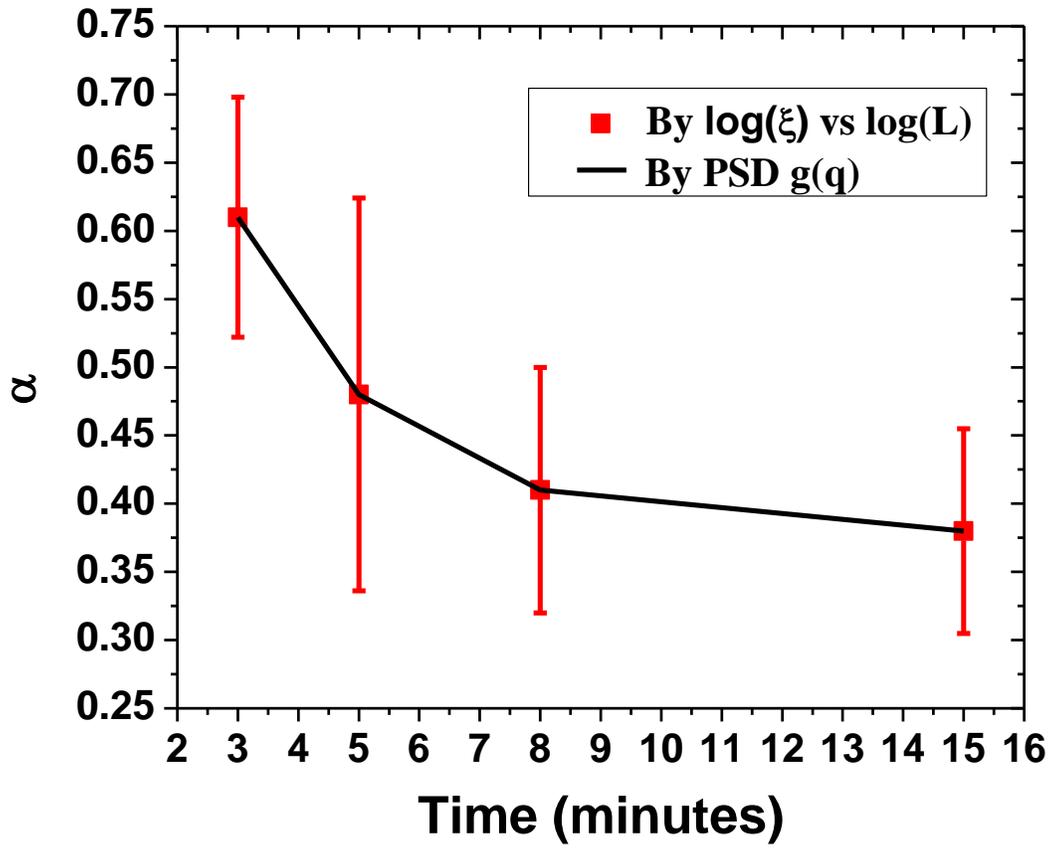

**Figure. 5**

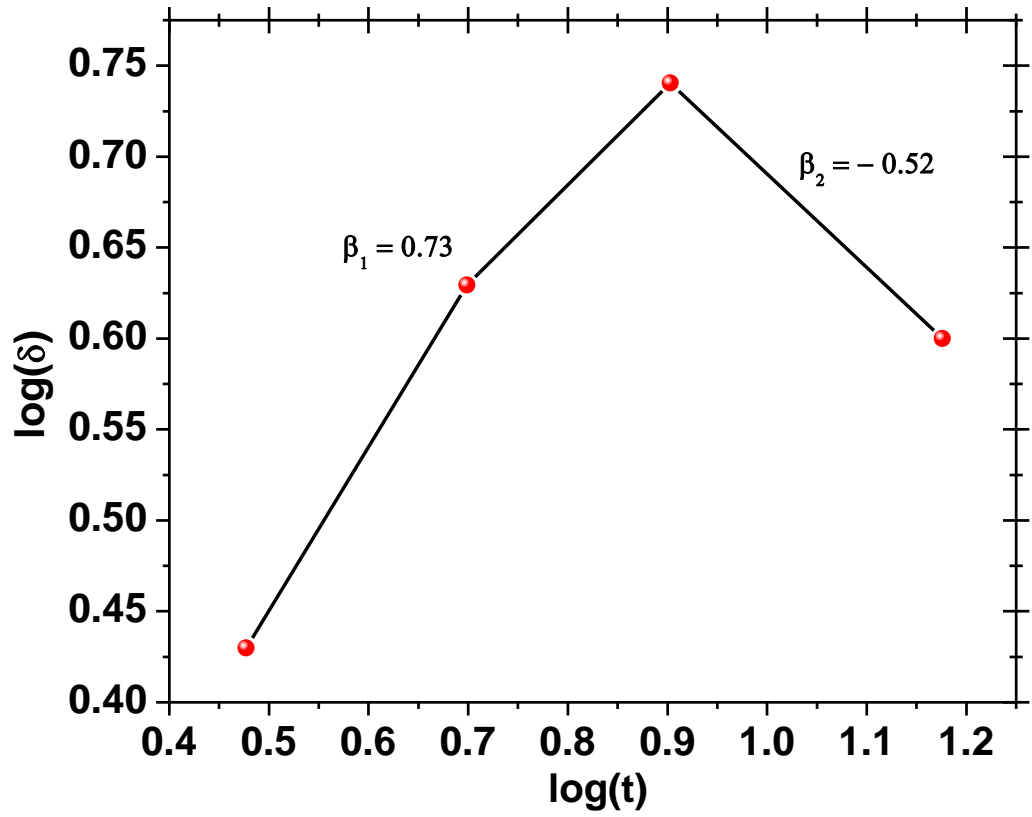

Figure. 6

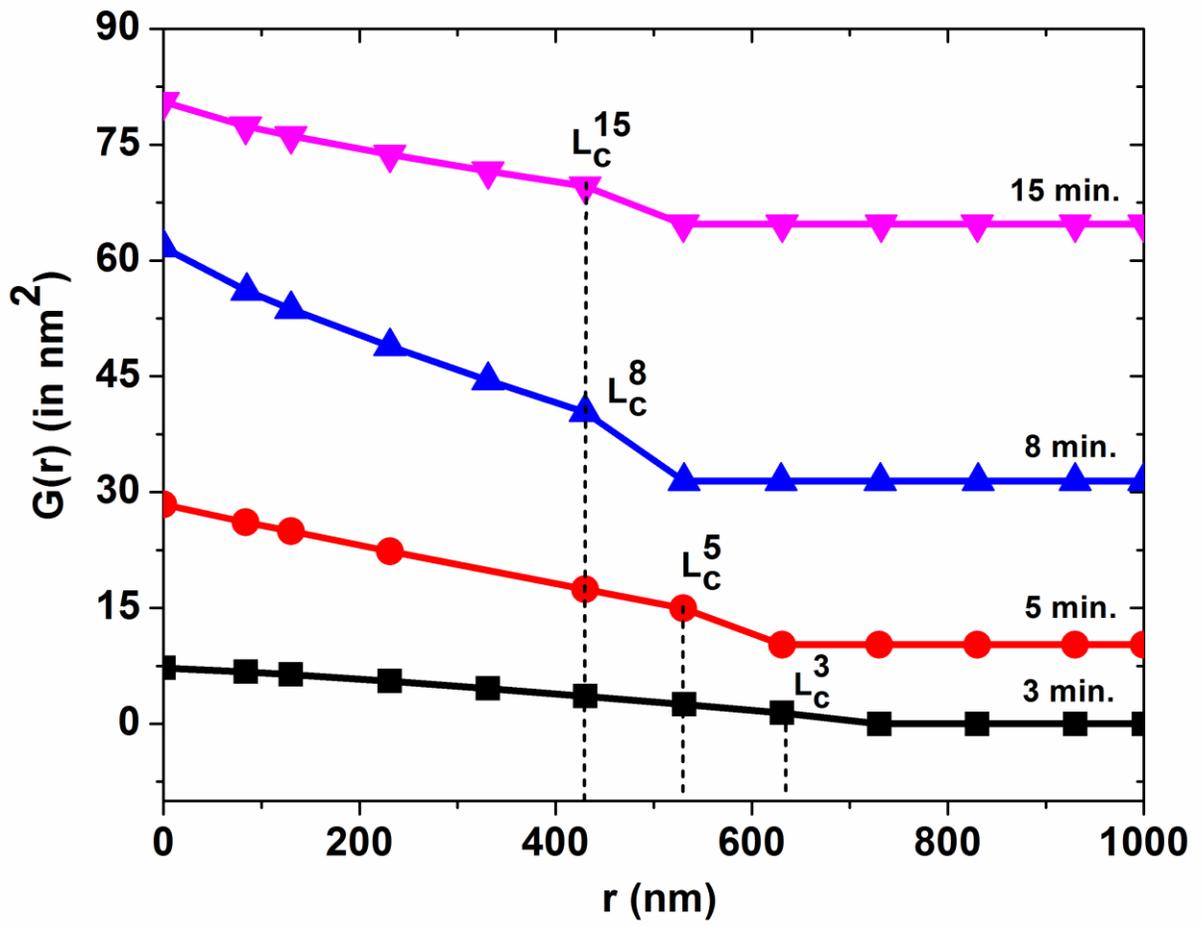

**Figure. 7**

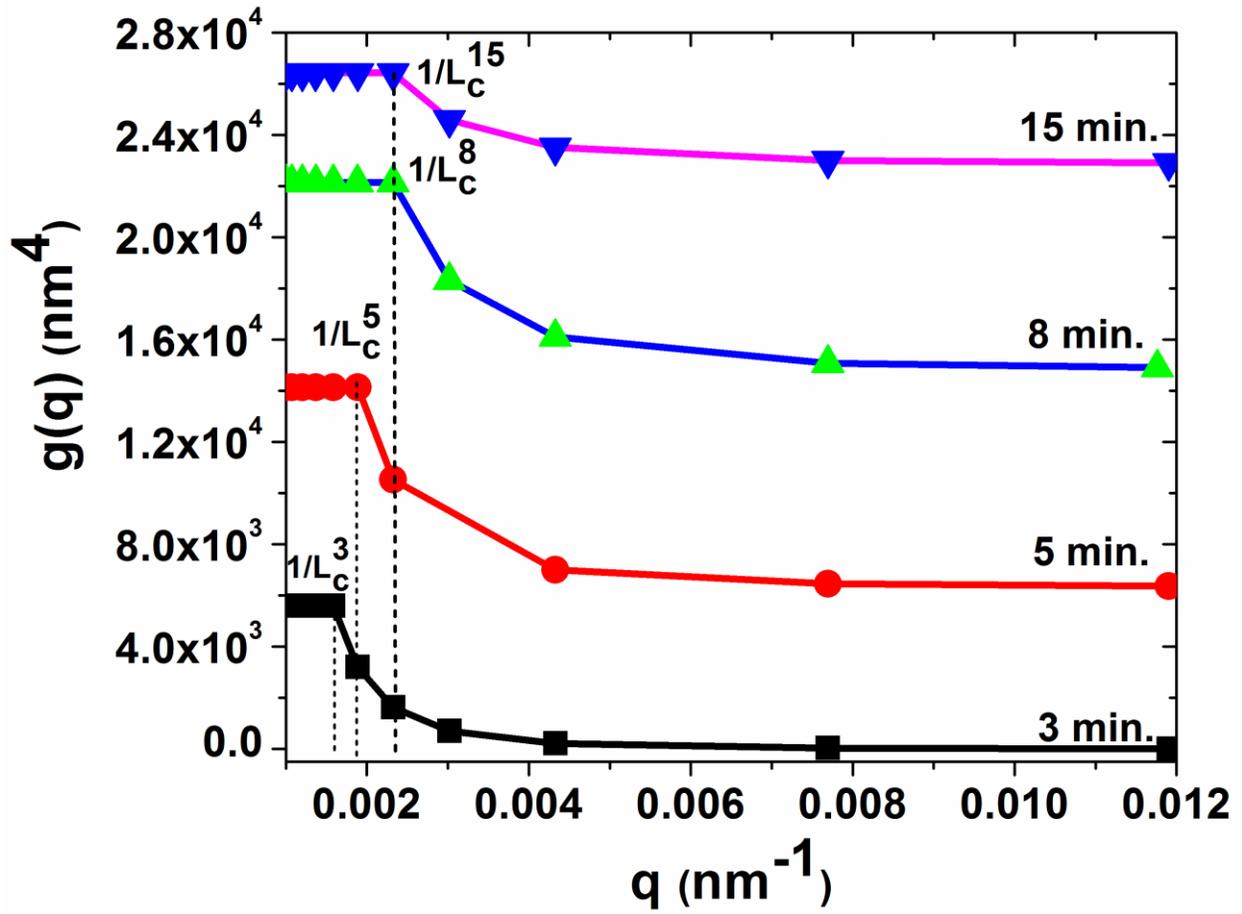

**Figure. 8**

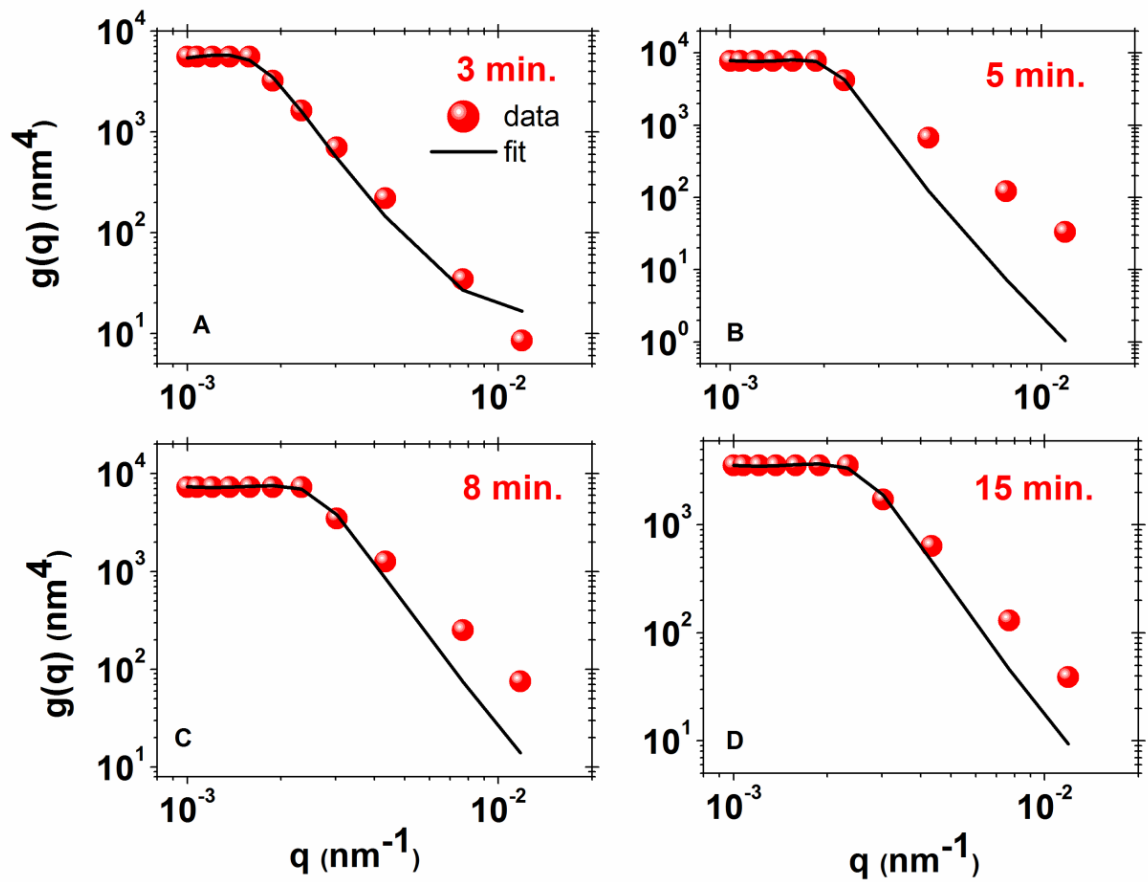